\documentclass[11pt]{article}
\usepackage{moriond,epsfig}

\def\Journal#1#2#3#4{{#1} {\bf #2}, #3 (#4)}

\def\PLB{{\em Phys. Lett.}  B}
\def\PRL{\em Phys. Rev. Lett.}
\def\PRD{{\em Phys. Rev.} D}

\def\be{\begin{equation}}
\def\ee{\end{equation}}
\def\bea{\begin{eqnarray}}
\def\eea{\end{eqnarray}}

\begin{document}
\vspace*{4cm}
\title{SOLAR SYSTEM DARK MATTER}

\author{Stephen L. Adler}

\address{Institute for Advanced Study, Einstein Drive, Princeton NJ 08540, USA}

\maketitle\abstracts{I review constraints on solar system-bound dark
matter, and discuss the possibility that dark matter could be
gravitationally bound to the earth and other planets.  I briefly
survey various empirical constraints on such planet-bound dark
matter, and discuss effects it could produce if present, including
anomalous planetary heating and flyby velocity changes.}

\section{Gravitationally Bound Dark Matter}

Cosmology suggests that only around 4\% of the mass-energy density
of the universe is ordinary baryonic matter.  In addition to
ordinary matter, around 23\% of the mass-energy density consists of
gravitationally attractive ``dark matter'', and the remaining 73\%
consists of gravitationally repulsive ``dark energy''.

Little is known about dark matter, other than that it is
electrically neutral -- hence dark, since it does not radiate
photons -- and that it interacts gravitationally through its
mass-energy density.  Among the unanswered questions are:  Is dark
matter bosonic or fermionic?   Is it self-annihilating (i.e.,  its
own antiparticle, or having equal abundances of particles and
antiparticles).  What is the mass (or are the masses, if more than
one constituent) of dark matter particles?  What are the
non-gravitational interactions of dark matter?

Dark matter can be gravitationally bound on different scales.  We
enumerate possibilities in the following subsections.

\subsection{Galactic Halo Dark Matter}\label{subsec:gal}

Our galaxy is believed to be surrounded by a dark matter halo, with
a mass density (to within a factor of about two) of $\rho \sim 0.3
{\rm GeV}/c^2 {\rm cm}^{-3}$. The galactic dark matter is believed
to have a roughly Maxwellian velocity distribution of the form
\begin{equation}\label{vel}
f \propto v^2 e^{-3v^2/(2 {\bar v}^2)},~~~~~ \bar v \sim 270 {\rm km}\,{\rm s}^{-1}~~~.
\end{equation}
In this formula, $v=w_{\rm dm}+v_{\oplus}$, with $w_{\rm dm}$ the
velocity of the galactic dark matter relative to the earth, 
and with $v_{\oplus}$ the velocity of the earth through the
galaxy.  The latter has two components, the earth's orbital velocity
(to which, at a low level, one should add the surface rotational
velocity), and the velocity of the solar system barycenter through
the galaxy. During part of the year these velocities add, and during
part they subtract, leading to an annual modulation of the dark matter detection rate 
observed from earth which was proposed by Drukier, Freese, and
Spergel \cite{druk} and Freese, Frieman, and Gould \cite {freese}
as a signal for searching for dark matter. Possible observation of
galactic dark matter through this signal has been reported by the
DAMA/LIBRA experiment \cite{dama}.

\subsection{Solar System-Bound Dark Matter?}

In addition to dark matter bound to the galactic center of mass, it
is possible that dark matter could be bound on smaller scales, such
as to the sun or to planets.  Dark matter bound to the sun would
change planetary orbits, by giving additional perihelion precessions
beyond that predicted by general relativity, and by changing the
Kepler's third law relation between the period and semi-major axis
of orbits.  A number of authors (see Fr\`ere, Ling and Vertongen
\cite{frere}, Sereno and Jetzer \cite{sereno}, Iorio \cite{iorio},
and Khriplovich and Pitjeva \cite{khrip}) have studied these
effects, with the conclusion that the density of sun-bound dark
matter is constrained by $\rho < 10^5 {\rm GeV}/c^2 {\rm cm}^{-3}$.
Because the earth's rotational velocity can add to the earth's
orbital velocity, or subtract from it, depending on the sidereal
time, sun-bound dark matter at densities well above the galactic
density could be detected through a search for a daily sidereal time
modulation in the DAMA/LIBRA and similar experiments.  Such a
modulation would have a 24 hour period, which would allow it to be
distinguished from detector channelling effects of galactic halo dark
matter, which as discussed in Avignone, Creswick and Nussinov
\cite{avignone} would lead to a sidereal time modulation with a 12
hour period.

\subsection{Earth-Bound  (or Planet-Bound) Dark Matter?}

Another possibility is that there may be dark matter gravitationally
bound to the earth.  As shown in Adler \cite{adler}, one can place a
direct bound on this, by using the fact that for a satellite of
negligible mass in a circular orbit around a body of mass $M$,
tracking to give the orbit radius $R$ and orbit period $T$ give the
product $GM$ of the Newton gravitational constant and the mass,
\begin{equation}\label{gmformula}
GM=4\pi^2R^3/T^2~~~~.
\end{equation}
This formula, and its generalizations to include elliptical orbits
and perturbations on an inverse square law force, can be applied as
follows to give a useful bound on dark matter orbiting the earth.
Consider first the LAGEOS geodetic satellite, a very dense nearly
spherical satellite in orbit at a radius $R \sim 12,300 {\rm km}$.
Accurate tracking of LAGEOS, and modelling of drag effects, gives an
accurate value for $GM_{\oplus}$, where $M_{\oplus}$ is the mass of
the earth plus the mass of any dark matter below the LAGEOS orbit.
Similarly, accurate tracking of lunar orbiters gives $GM_m$, with
$M_m$ the mass of the moon.  A more accurate way of getting the
moon's mass (times $G$) is through tracking of the Eros asteroid
flyby, with gives a very accurate figure for
$R_{\oplus/m}=GM_{\oplus} /GM_m $.  Finally, lunar laser ranging,
which measures the relative distance between the earth and the moon,
gives the product $GM_{\rm combined}$ for the earth-moon system,
including the mass $GM_{\rm dm}$ of dark matter in orbit between the
LAGEOS orbit and the moon's orbit,
\begin{equation}\label{combined}
GM_{\rm combined}=GM_{\oplus}+GM_m+GM_{\rm dm}  ~~~~.
\end{equation}
Using this, and the earth to moon mass ratio obtained from the Eros
flyby, we get
\begin{equation}\label{subtract}
GM_{\rm dm} \simeq GM_{\rm combined}-GM_{\oplus}-GM_{\oplus}/R_{\oplus/m}~~~,
\end{equation}
where I have used an approximately equals sign because there could
be a roughly 1\%  correction coming from dark matter in the
vicinity of the moon or the Eros asteroid (see \cite{adler} for
details). Using the best current numbers (supplied to me by Slava
Turyshev \cite{slava}) one finds from equation (\ref{subtract}) that
\begin{equation}\label{result}
GM_{\rm dm} \simeq (0.3 \pm 4) \times 10^{-9} GM_{\oplus}~~~,
\end{equation}
with the dominant error coming from lunar ranging (which will be
considerably improved over the next decade).  {\it If} this bound
were attained, and if the dark matter mass were uniformly
distributed below the moon's orbit, the density would be $\rho \sim
6 \times 10^{10} {\rm GeV}/c^2 {\rm cm}^{-3}$, which is a much
higher density than both the galactic halo dark matter density and 
the upper bound on the density of sun-bound dark matter.  So in
principle, there could be substantial amounts of dark matter
gravitationally bound to the earth (or generalizing from this, to
other planets).

\section{Possible Applications of Earth- and Planet-Bound Dark Matter}

What I now want to say is necessarily speculative, since it is conditioned 
on the possibility that there may be significant amounts of dark matter 
gravitationally bound to the earth and other planets.  

\subsection{Jovian Planet Anomalies}  

In Adler \cite {adler1} I suggest that dark matter accretion may play a 
role in explaining certain features of the Jovian planets Jupiter, Saturn, 
Uranus, and Neptune.  From de Pater and Lissauer \cite{depater} one learns 
that the surface heat fluxes $H$ of these planets, in ${\rm ergs}/{\rm cm}^2{\rm s}$, 
are  5440 (Jupiter), 2010 (Saturn), $<$42 (Uranus), and 433 (Neptune).  The larger heat 
fluxes have not been completely accounted for (for example, through internal radioactive 
decays) and it is possible that a new mechanism may be at work.  In \cite{adler1} I suggest 
that the accretion of planet-bound dark matter may account for unexplained internal heat 
production. This would be possible if each planet has an associated dark matter cloud (perhaps linked, 
as suggested in \cite{frere}, to formation of the planet), and if the dark matter is accreted  
with a low energy release efficiency, so that 
bounds on earth heat production arising from accretion of galactic dark matter are not 
violated.  Another peculiar feature of the Jovian planets is that Uranus, which is otherwise 
very similar to Neptune, has a much lower internal heat production, as well as being the only 
planet which has its rotation axis lying on its side rather than nearly normal to the ecliptic.  
I suggest that the collision that is hypothesized to have tilted the axis of Uranus could also have 
knocked Uranus out of its associated dark matter cloud, accounting also for the anomalously low 
internal heat production.  

\subsection{Flyby Anomaly}

When spacecraft are put into near-earth hyperbolic orbit flyby trajectories, there is a several hour segment 
when the spacecraft is close to the earth and cannot be tracked.  When the incoming orbit is extrapolated, using 
the best orbit fitting programs, to predict the outgoing orbit, a discrepancy between the predicted and observed 
outgoing velocity is found, as reported in Anderson et al. \cite{anderson}  For example, the Galileo II flyby 
on 12/8/92 shows a velocity discrepancy of -4.6 mm/s (with an estimated error of 1 mm/s), that is the outgoing velocity is lower than expected, while 
the Near flyby on 1/23/98 shows a velocity discrepancy of 13.46 mm/s (with an estimated error of only .01 mm/s), that is the 
outgoing velocity is {\it higher} than expected.  The discrepancies are roughly of order $10^{-6}$ of the total velocity, so 
constitute a relatively large effect, and as we have just seen, can have either sign.  

There are at least four possibilities for explaining these flyby anomalies: 

\noindent i.  The effect is an {\bf artifact}, resulting from the omission from the orbit fitting programs 
of known physics.  However, no one has been able to identify a candidate so far, since all the obvious 
things, including  moon and planetary perturbations, special and general relativistic corrections, solar wind effects, Van Allen belt collisions,  tides, and thermal effects, have been taken into account, or estimated and shown to be too small 
to be relevant.  

\noindent ii.   New {\bf electromagnetic physics} is involved.  However, this seems unlikely, since experimental tests of 
quantum electrodynamics and vacuum linearity have a much higher accuracy than 1 part in $10^6$. 

\noindent iii.  New {\bf gravitational} physics is involved.  This would have to be quite unconventional.  MOND (modified Newtonian 
dynamics), which suggests a modification in Newtonian gravitation for very low accelerations, does not predict the flyby 
anomalies.  I (and others) have taken a look at the parameterized post-Newtonian (PPN) formalism that gives a phenomenology for 
studying  (in the weak gravitational field, low velocity regime)    metric theories of gravitation that differ from general relativity.  Using the modified equation of motion given in Will \cite{will} and the 2006 bounds on the PPN parameters, I 
find that all terms are too small, by several orders of magnitude, to account for the flyby anomaly.  This still leaves the 
possibility of non-metric gravitational theories, and some ideas for these have been proposed \cite{nonmetric}.  

\noindent iv.   The effect comes from {\bf collisions of the flybys with earth-bound dark matter}  This possibility is analyzed 
in my paper Adler \cite{adler2}.  Velocity decreases can arise from ordinary drag, as a result of elastic scattering 
$D(m_D) + N \to D(m_D) + N$, with $m_D$ the mass of a dark matter particle $D$ and $N$ a spacecraft nucleon.  Velocity increases 
can arise from exothermic inelastic scattering $D(m_D) + N \to D^{\prime}(m_{D^{\prime}}) +N$, with the primary dark matter 
particle of mass $m_D$ scattering into a secondary of mass $m_{D^{\prime}} < m_D$, and with the secondary recoiling predominantly 
backward, so that the nucleon gets a forward kick.

My analysis of the dark matter possibility shows that the following strong constraints are required, in order for dark 
matter collisions to provide an explanation for the flyby anomaly:   

\noindent i.  The dark matter in orbit around the earth must be localized well within the moon's orbit 
and not too near the earth.

\noindent ii.  The dark matter mass must be much smaller than a GeV. 

\noindent iii.   The dark matter scattering cross section on nucleons must be high, of order 
$10^{-33}$ to $10^{-27} ~ {\rm cm}^2$.  

\noindent iv. The dark matter must be non-self-annihilating and stable in the absence of nucleons.  

These constraints require a form of dark matter that is very different from the customary assumption that dark matter is a multi-GeV lightest supersymmetric particle.  

My current work consists of modelling dark matter orbiting the earth, with two species in different configurations, one 
of which elastically scatters from nucleons, and one of which inelastically scatters, to try to fit the Anderson group 
data.  I assume dark matter distributions obtained by averaging a circular orbit of radius $r$ at an angle $\chi$ relative to the earth's rotation axis, over precession around the earth's axis, giving a truncated spherical shell of radius $r$,  
height $2|z| = 2 r \sin \chi$, and relative density $1/\sqrt{(r \sin \chi)^2 - z^2}$.  For a given point 
characterized by  $r,z,\chi$ there 
are two relevant dark matter velocities that correspond respectively to an up-moving and a  down-moving dark matter orbit segment 
passing through the point, and the corresponding velocity vectors are easily worked out. From these velocities and the spacecraft 
velocity, one can calculate the cross section-averaged velocity change imparted to a spacecraft nucleon, and integrating the corresponding work 
increment  over the spacecraft orbit gives the final energy and velocity change.  I currently have the core programs for this calculation written, and 
now have to do a systematic search of the parameter space to try to find a reasonable fit to the data.

\section*{Acknowledgments}
This work was supported in part by the U. S. Department of Energy under grant No. DE-FG02-90ER40542.


\section*{References}
\begin{thebibliography}{99}


\bibitem{druk} A. K. Drukier, K. Freese and D. N. Spergel, \Journal{\PRD} {33} {3495} {1986}.

\bibitem{freese} K. Freese, J. Frieman and A. Gould, \Journal{\PRD} {37} {3388} {1988}.

\bibitem{dama} R. Bernabei {\it et al}, {\em Eur. Phys. J.} C {\bf 56}, 333 (2008).

\bibitem{frere} J.-M. Fr\`ere, F.-S. Ling and G. Vertongen, \Journal{\PRD} {77}{083005} {2008}.

\bibitem{sereno} M. Sereno and Ph. Jetzer, {\it Mon. Not. R. Astron. Soc.} {\bf 371}, {626} (2006). 

\bibitem{iorio} L. Iorio, {\it J. Cosmol. Astropart. Phys.} JCAP05, 002 (2006). 

\bibitem{khrip} I. B. Khriplovich and E. V. Pitjeva, {\it Int. J. Mod. Phys.} D {\bf 15}, 615 (2006); 
  I. B. Khriplovich, {\it Int. J. Mod. Phys.} D {\bf 16}, 1475 (2007). 
  
\bibitem{avignone} F. T. Avignone III, R. J. Creswick and S. Nussinov, arXiv:0807.3758.  

\bibitem{adler} S. L. Adler, {\it J. Phys. A: Math. Theor} {\bf 41}, 412002 (2008). 

\bibitem{slava} S. Turyshev, private communication (2008).  

\bibitem{adler1} S. L. Adler, \Journal{\PLB} {671} {203} {2009}. 

\bibitem{depater} I. de Pater and J. J. Lissauer, Planetary Sciences, Cambridge University Press (2001), 
pp. 224-225. 

\bibitem{anderson}  J. D. Anderson {\it et al}, \Journal{\PRL} {100} {091102} {2008}. 

\bibitem{will} C. M. Will, Theory and Experiment in Gravitational Physics, Revised Edition, Cambridge University 
Press (1993), Sec. 6.2. 

\bibitem{nonmetric}  M. E. McCulloch, arXiv:0806.4159; M. B. Gerrard and T. J. Sumner, arXiv:0807.3158.  

\bibitem{adler2} S. L. Adler, \Journal{\PRD} {79} {023505} {2009}.  





\end{thebibliography}
\end{document}